\documentclass[12pt,epsf]{article}
\newlength{\abstractwidth}
\usepackage[dvips]{graphicx}
\usepackage{subfigure}

\renewcommand{\thefootnote}{\fnsymbol{footnote}}
\renewcommand{\thanks}[1]{\footnote{#1}} 
 \newcommand{\starttext}{
\setcounter{footnote}{0}
\renewcommand{\thefootnote}{\arabic{footnote}}}
\renewcommand{\theequation}{\thesection.\arabic{equation}}
\newcommand{\be}{\begin{equation}}
\newcommand{\bea}{\begin{eqnarray}}
\newcommand{\eea}{\end{eqnarray}}
\newcommand{\beq}{\begin{equation}}
\newcommand{\ee}{\end{equation}}
\newcommand{\eeq}{\end{equation}}

\newcommand{\<}{\langle}

\renewcommand{\>}{\rangle}
\def\ba{\begin{eqnarray}}
\def\ea{\end{eqnarray}}

\def\14{{1\over4}}
\def\12{{1 \over 2}}

\def\h3{h^{3\over 2}}

\def\>{\rangle}
\def\<{\langle}

\def\0cc{$\Lambda = 0$}

\begin{document}
\renewcommand{\theequation}{\thesection.\arabic{equation}}
\begin{titlepage}
\bigskip
\rightline{} \rightline{}

\bigskip\bigskip\bigskip\bigskip

\centerline{\Large \bf {Cosmic Natural Selection }}

\bigskip\bigskip
\bigskip\bigskip

\centerline{\it  L. Susskind  }
\medskip
\centerline{Department of Physics} \centerline{Stanford
University} \centerline{Stanford, CA 94305-4060}
\medskip
\medskip

\bigskip\bigskip
\begin{abstract}
I make a number of comments about Smolin's theory of Cosmic
Natural Selection.
\end{abstract}

\end{titlepage}
\starttext \baselineskip=18pt \setcounter{footnote}{0}

In an unpublished note I criticized Smolin's theory of
cosmological natural selection \cite{sm}which argues that we live
in the fittest of all possible universes. By fitness, Smolin means
the ability to reproduce. In my criticism I used the example of
eternal inflation which is an extremely efficient reproduction
mechanism. If Smolin's logic is applied to that example it would
lead to the prediction that we live in the universe with the
maximum cosmological constant. This is clearly not so.

Smolin proposes that the true mechanism for reproduction is a
bouncing black hole singularity that leads to a new universe
behind the horizon of  every black hole. Thus Smolin suggests that
the laws of nature are determined by maximizing the number of
black holes in a universe.

Smolin also argues that it is not obviously wrong that our
physical parameters, including the smallness of the cosmological
constant, maximize the black hole formation. To make sense of this
idea, one must assume that there is a very dense discretuum of
possibilities, in other words a rich landscape of the kind that
string theory suggests \cite{bp}\cite{kklt}\cite{ls}\cite{ad}.

The detailed astrophysics that goes into Smolin's estimates in
extremely complicated--too complicated for me--but the basic
theoretical assumptions that go into the theory can be evaluated,
especially in light of what string theory has taught us about the
landscape and about black holes.

As I said, there are two mechanisms, eternal inflation and black
hole production that can contribute to reproduction, and it is
important for Smolin's scenario that black holes dominate.
Considering the low density of black holes in our universe and the
incredible efficiency of exponential inflation, it seems very hard
to believe that black holes win unless eternal inflation is not
possible for some reason.

Smolin's argues that we know almost nothing about eternal
inflation but we know a great deal about black holes including the
fact that they really exist. This is a bit disingenuous. Despite a
great deal of serious effort \cite{H} \cite{Fid}, the thing we
understand least is the resolution of black hole and cosmic
singularities. By contrast, eternal inflation in a false vacuum is
based only on classical gravity and semiclassical Coleman de
Luccia bubble nucleation \cite{Coleman}\cite{guth}.

The issue here is not whether the usual phenomenological inflation
was of the eternal kind although that is relevant. Eternal
inflation taking place in any false vacuum minimum on the
landscape would favor (in Smolin's sense) the maximum cosmological
constant. But for the sake of argument I will agree to ignore
eternal inflation as a reproduction mechanism.

The question of how many black holes are formed is somewhat
ambiguous. What if two black holes coalesce to form a single one.
Does that count as one black hole or two? Strictly speaking, given
that black holes are defined by the global geometry, it is only
one black hole. What happens if all the stars in the galaxy
eventually fall into the central black hole? That severely
diminishes the counting. So we better assume that the bigger the
black hole, the more babies it will have. Perhaps one huge black
hole spawns more offspring that $10^{22}$ stellar black holes.

That raises the question of what exactly is a black hole? One of
the deepest lessons that we have learned over the past decade is
that there is no fundamental difference between elementary
particles and black holes. As repeatedly emphasized by 't Hooft
\cite{'tHooft}\cite{spec}\cite{Horowitz:1996nw}, black holes are
the natural extension of the elementary particle spectrum. This is
especially clear in string theory where black holes are simply
highly excited string states. Does that mean that we should count
every particle as a black hole?

Smolin's theory requires not only that black hole singularities
bounce but that the parameters such as the cosmological constant
suffer only very small changes at the bounce. This I find not
credible for a number of reasons. The discretuum of string theory
does indeed allow a very dense spectrum of cosmological constants
but neighboring vacua on the landscape do not generally have close
values of the vacuum energy. A valley is typically surrounded by
high mountains, and neighboring valleys are not expected  to have
similar energies.

Next--the energy density at the bounce is presumably Planckian.
Supposing that a bounce makes sense, the new universe starts with
Planckian energy density. On the other hand Smolin wants the final
value of the vacuum energy density to be very close to the
original. It sounds to me like rolling a bowling ball up to the
top of a very high mountain and expecting it to roll down, not to
the original valley, but to one out of $10^{120}$ with almost
identical energy. I find that unlikely.

Finally, we have learned some things about black holes over the
last decade that even Stephen Hawking agrees with \cite{nyt}.
Black holes do not lose information. The implication \cite{hm} is
that if there is any kind of universe creation in the interior of
the black hole, the quantum state of the offspring is completely
unique and can have no memory of the initial state. That would
preclude the kind of slow mutation rate envisioned by Smolin.

Smolin seems to think that there is significant evidence that
singularity resolution (by bounce) is imminent. Loop quantum
gravity, according to him, is on the threshold of accomplishing
this. Perhaps it will. But either it will be consistent with
information conservation in which case the baby can have no memory
of the parent, or it will not. If not it probably means that Loop
gravity is inconsistent.


\end{document}